\documentclass[runningheads]{svmult}

\usepackage{makeidx}   
\usepackage{graphicx}  
\usepackage{subeqnar}  
\usepackage{multicol}  
\usepackage{physprbb}  
\makeindex             

\begin{document}

\title*{ GRBs, Fireballs and Precessing Gamma Jets}
\toctitle{GRBs and Fireball \protect\newline versus Precessing
Gamma Jets}
%
%
\titlerunning{GRBs, Fireballs and Precessing Gamma Jets}
%
\author{Daniele Fargion \inst{1,2}
}

\authorrunning{Daniele Fargion}
%
%


\institute{Universit\'{a} degli Studi di Roma I, $La Sapienza ^1$
and
$INFN ^2$,\\
 Piazzale Aldo Moro 5 , 00185,Rome,Italy}

\maketitle              

\begin{abstract}

Fireballs are  huge isotropic explosions models widely believed
 to explain Gamma Ray Burst, GRBs (Piran,1999); ever-new versions consider
  wide beamed ($10^o$) Jet explosions hitting external shells.
  On the contrary, since 1994-1998, we  argued (Fargion
  1995-2000; see also Blackman et all.1996)
 that GRBs (as well as Soft Gamma Repeaters SGR)
are spinning and precessing Gamma Jets, produced by collimated
$e^+$,$e^- $ Jet via Inverse Compton Scattering, in a very narrow
($0.1^o$) angles, blazing and flashing the observer. The Jet
arises in Super-Nova (SN) explosions; its energy decays slowly
 from earliest SN powers (corresponding to GRB) toward
lower stable power as Soft Gamma Repeaters (SGR) regimes. GRBs
and SGRs shared (sometimes) same spectra and time structure: then
SGRs are low-power GRBs, but without
 SN relics (or GRB afterglows, signatures of
Jets in SN-GRBs). Moreover weak isolated X-ray precursor
signals,(such as $GRB980519$, $GRB981226$, $GRB000131$),
corresponding to huge isotropic  $\sim 10^{47}$ erg $s^{-1}$,
followed by the extreme GRB  $\sim {10^{52}}$ erg $s^{-1}$ powers,
disagree with any Fireball explosive scenarios. We naturally
interpret these X-Ray precursors as rare earliest marginal blazes
of outlying X conical precessing Jet tails, surrounding the
$\gamma$ Jet, later hitting in-axis as a GRB.
\end{abstract}

\begin{figure}[h]
\begin{center}
 \includegraphics[width=6 cm , height=6 cm]{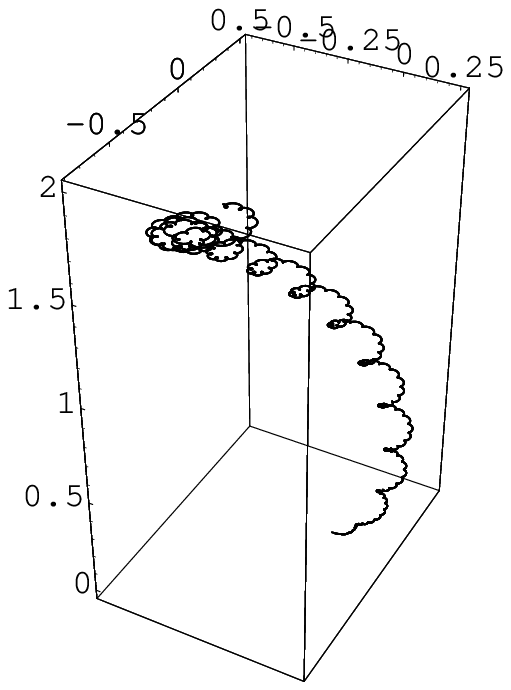}
 \includegraphics[width=6 cm , height=6 cm]{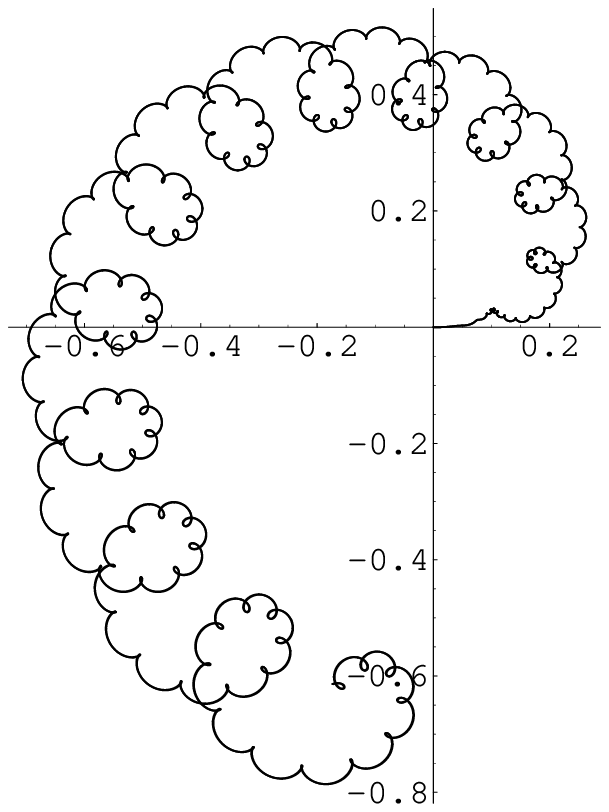}

 \end{center}
 \caption{Left: Three dimensional space evolution of a Spinning while Precessing
  X -$\gamma$ Jet, leading by its blazing  to X precursor and to the main $\gamma$ GRB.
   Right: The same jet pattern observed from above
  along the vertical axis , on a two dimensional plane.}
 \label{eps4}
\end{figure}


 Since $GRB980425$ we argued (Fargion 1998-2000)
  that GRBs and SGRs can be explained by a comprehensive theory
  where  a thin (tens of seconds) $\gamma$ beam
  Jet,  spinning in multi-precession, is sprayed by a Neutron Star, NS, or a Black Hole, BH,
  flashing and blazing the observer.  Indeed the extreme energy released
  in GRB990123 and GRB000131, ($\gg$ $10^{54}$  $erg $),
  (or even  twice as much, keeping into account neutrino  budget)
   leads to a  conflict with any isotropic GRB
   model: Schwarzschild  scale times
   (corresponding to the needed solar masses ), above milliseconds,
    disagree with the observed GRBs fine time structures (below a fraction of millisecond).
   GRBs and SGRs share, in a few cases, the same spectra
    (Fargion 1998-1999-2000;Wood et all 1999) and time structure, suggesting an unique
   model. The $\gamma$ Jet for GRB and SGR is produced,
   trough  Inverse Compton Scattering (ICS),
   by GeVs $e^+$ $e^-$ (secondaries of penetrating  GeVs $\mu^+$ $\mu^-$), scattering on  infrared photons,
     (Fargion,Salis 1995-1998), leading to a collimated, spinning and precessing  $\gamma$ (MeVs) precessing Jet.

\begin{figure}[h]
\begin{center}
\includegraphics[width=12 cm]{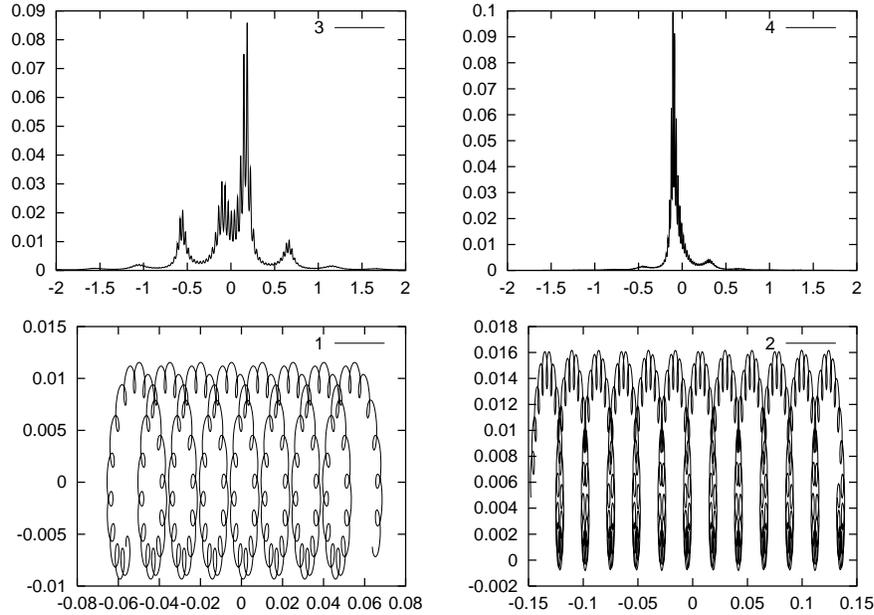}
 \end{center}
  \caption{ Lower figures show two different angular Jet
  patterns , as traced in  Fig.1.  Their bi-dimensional opening
  angles  while  Spinning and Precessing, are blazing  the observer at the center (origin
  ($0,0$))
  leading, by ICS, to the consequent GRB signal described above.
   Upper figures show the  the consequent X, GRB intensity  evolution (time in secs)
  derived by the ICS formula and the corresponding geometrical Jet patterns evolutions below.
   The X ray precursor may naturally arise in some pattern configurations.}
  \label{eps5}
\end{figure}


The peak $\gamma$ Jets has power of a Supernova ($10^{44} erg
s^{-1}$) appearing beamed as ($10^{52} erg s^{-1}$) decaying by
power law $\sim t^{-1}$  in 3-6 hour scale times, to ancient,
lower power SGRs stages. SGRs are powered by X-ray pulsars Jet
($10^{35} erg s^{-1}$) whose collimated beam is amplified up to
($10^{43} erg s^{-1}$). Both of GRB and SGR show an apparent
luminosity amplified by the  inverse of the beamed solid angle
($10^{7}- 10^{9}$). The  earliest and puzzling X-Ray precursors
in few GRBs (as well as SGRs) is an obvious peripherical off-axis
flashing followed by main in-axis GRB blaze, in some geometrical
configurations, as shown in simulations in Fig.2. Data on X-Ray
precursor and GRB are shown for comparison in Fig3.

\begin{figure}[h]
\begin{center}
 \includegraphics[width=.45\textwidth, bb=10 20 400 300]{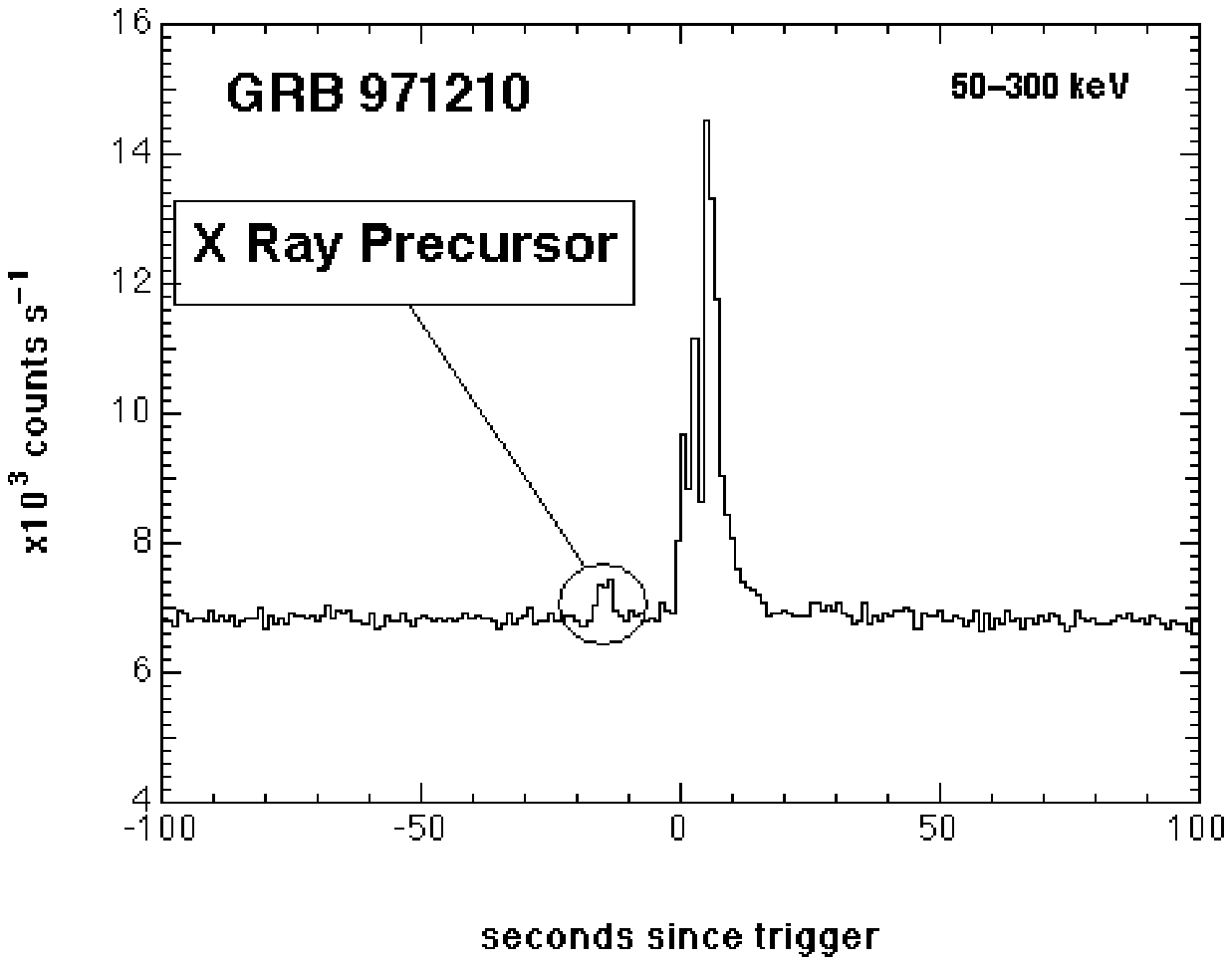}
 \includegraphics[width=6 cm, bb=10 20 400 300]{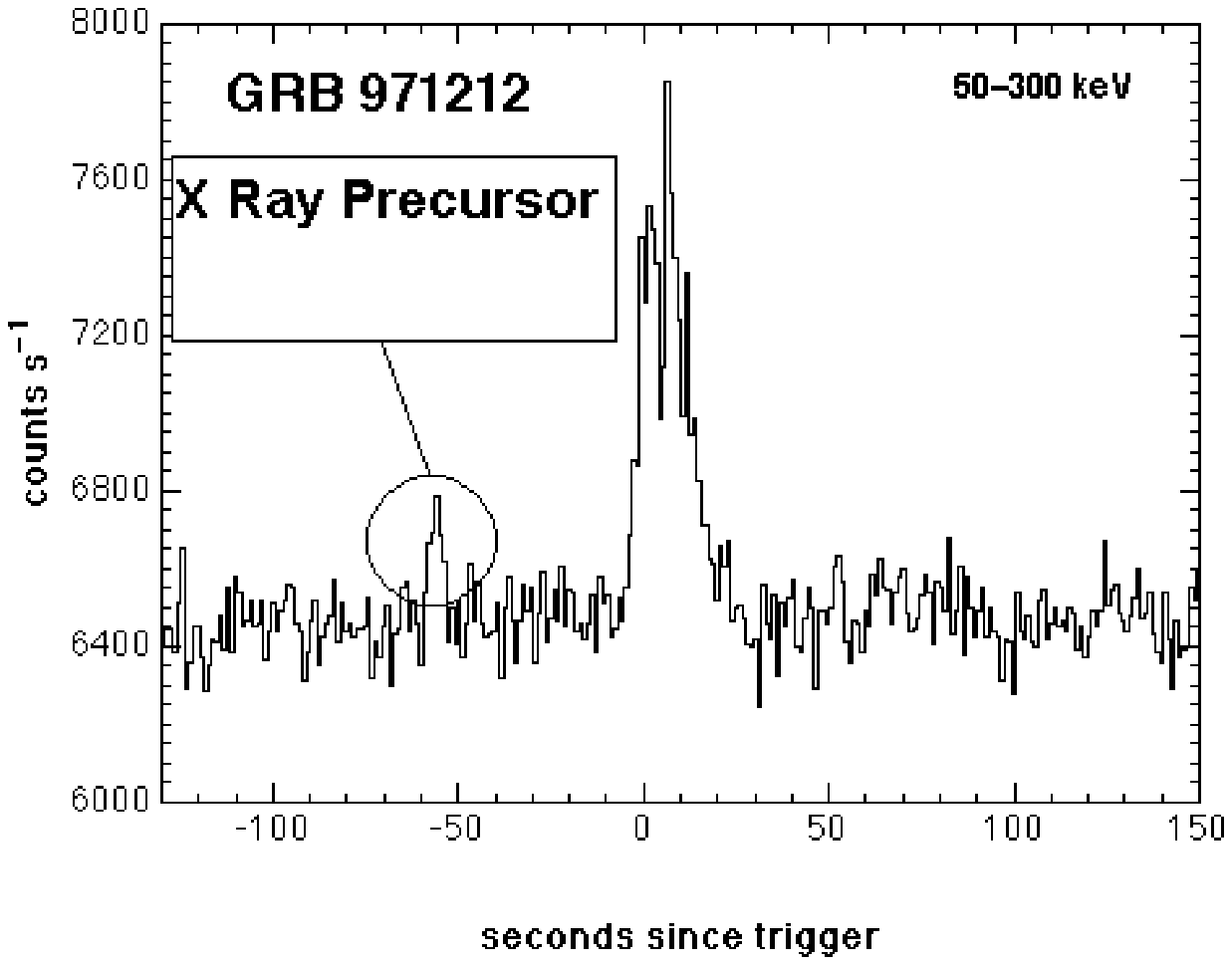}
\end{center}
\end{figure}
\vspace{-1.5cm}
\begin{figure}[h]
\begin{center}
 \includegraphics[width=6 cm, bb=10 20 400 370]{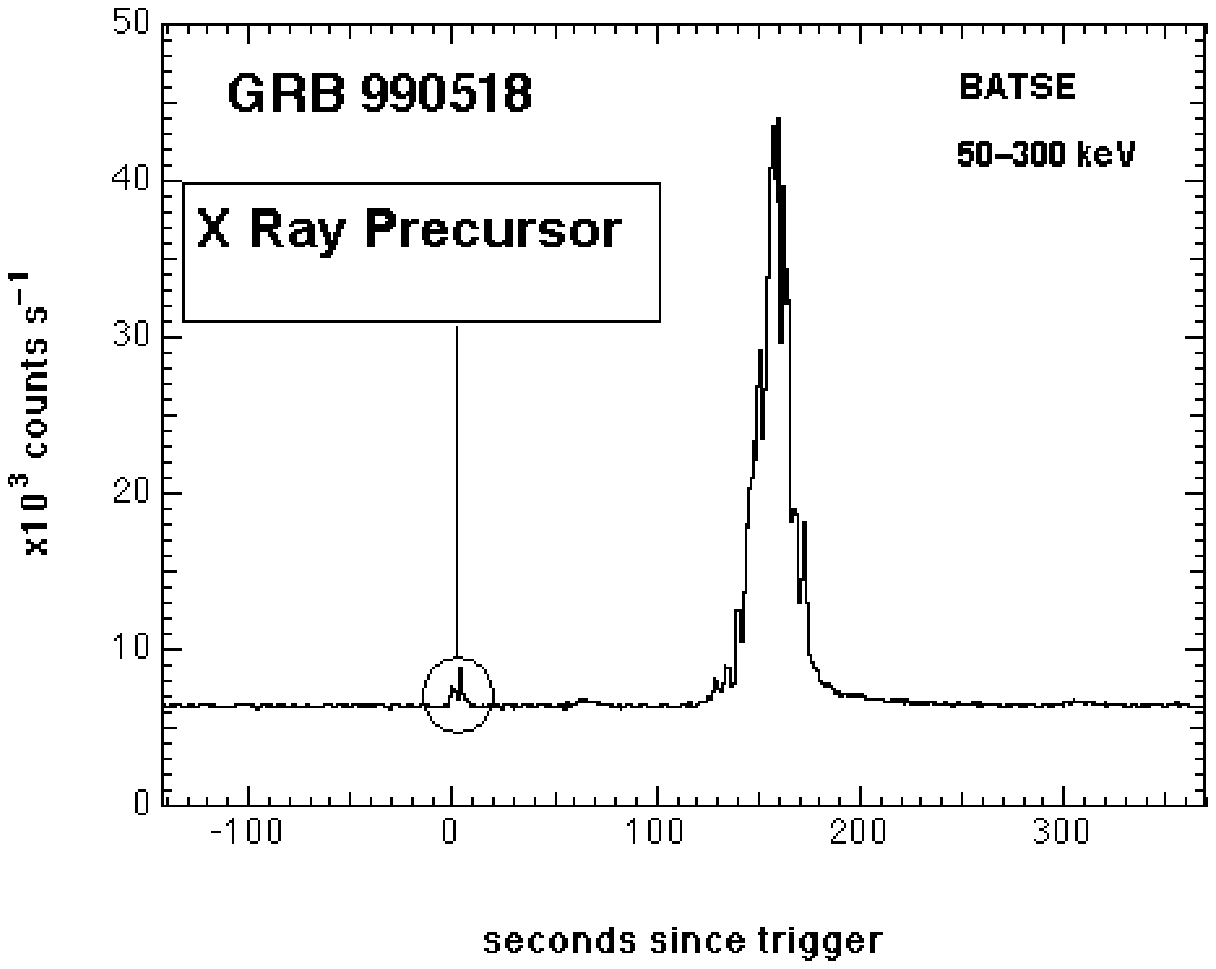}
\includegraphics[width=.435\textwidth]{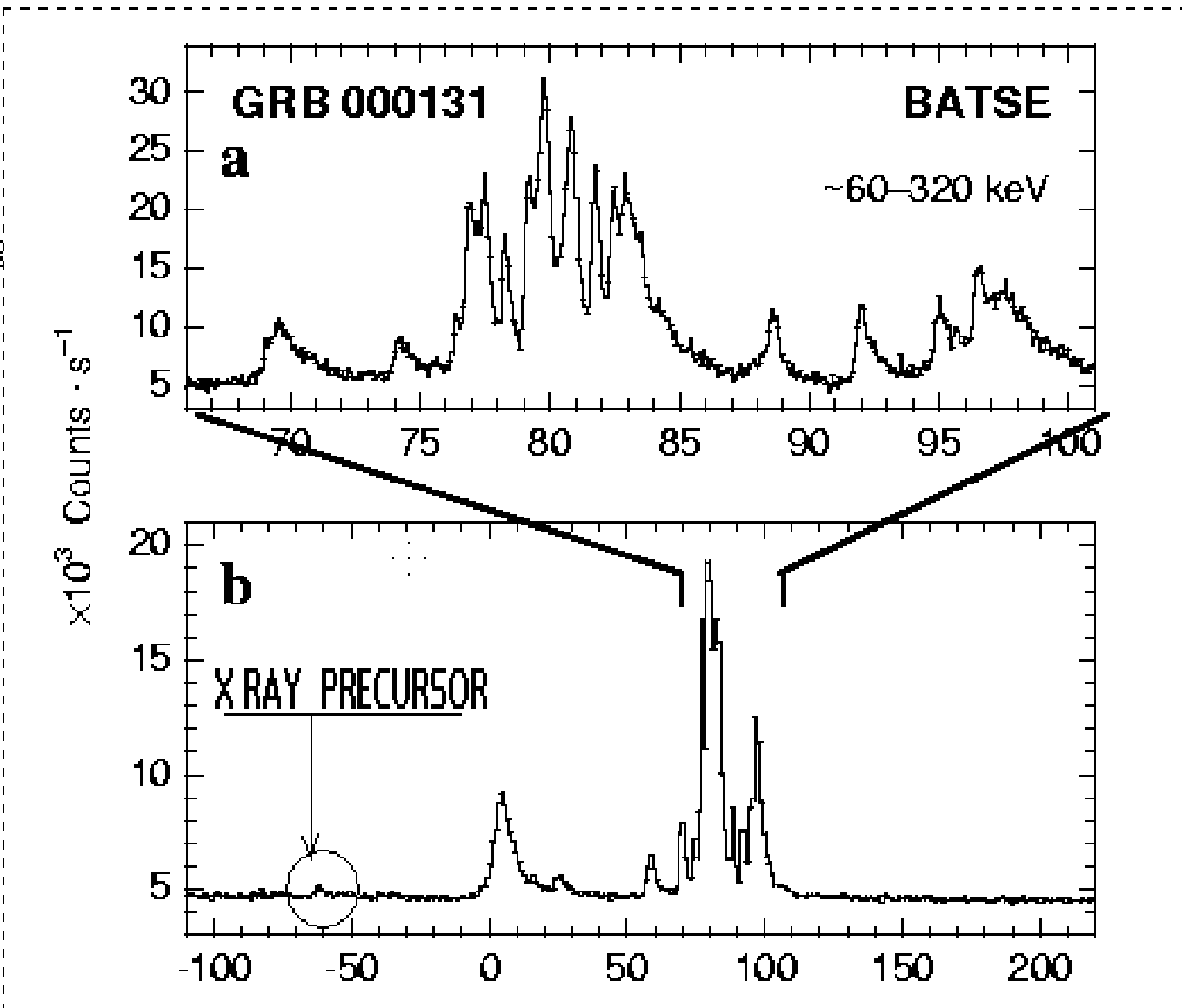}

\end{center}
\caption[]{Up: Time evolution and X precursors in GRB $971210$
and GRB $971212$. Down: The same evolution in GRB $990518$ and in
most distant (red-shift 4.5)  GRB $000131$; note the (surprising
for Fireball) tiny X-Ray precursor a minute before the main GRB.
\label{eps1}}
\end{figure}




%

\end{document}